\documentclass[aps,reprint]{revtex4-1}
\usepackage{graphicx}
\usepackage{amsmath}

\newcommand{\bmpi}{\ensuremath{\boldsymbol{\pi}}}

\begin{document}%
%
%
\title{Car-Parrinello Molecular Dynamics With A Sinusoidal Time-Dependent Potential Field}
\date{\today}
\author{Tobias Alznauer}
\author{Irmgard Frank}
\homepage{http://www.theochem.uni-hannover.de/}
\email{irmgard.frank@theochem.uni-hannover.de}
\affiliation{Theoretische Chemie\\ Leibniz Universit\"a{}t Hannover}
%
%
%

\begin{abstract}

We solve the problem of applying an external field in periodic boundary conditions
by choosing a sine potential.
We present an implementation in the Car-Parrinello molecular dynamics code (CPMD)
and discuss applications to electron and
ion transfers in complex molecular systems. \\

\end{abstract}

\maketitle

\section{Introduction}

Electrical fields may induce charge transport or also electrochemical reactions. 
To simulate such condensed-phase phenomena, it is desirable to have such an external field
implemented in a code which allows to perform Car-Parrinello dynamics \cite{Car1985}
using periodic boundary conditions \cite{cpmd}.
The theoretical treatment of external fields within Kohn-Sham theory \cite{Hohenberg1964,Kohn1965} is
in principle straightforward. The practical implementation in a code employing
periodic boundary conditions, however, leads to the question how potential
discontinuities of the potential energy at the borders of the unit cell should be
treated.
Several approaches to tackle this problem have been devised in the past years
\cite{Kudin2007a,Kudin2007b,Springborg2007,Springborg2008a,Springborg2008b,Kirtman2009}.
In the present paper we use a somewhat different approach with the aim
to devise an implementation in the CPMD code 
which is both conceptually simple and simple to use
in practical applications, even for time-dependent external fields.
We achieve this by representing an external field by
sine functions where the problem of discontinuities
does not arise.

\section{Theoretical background}

Car-Parrinello molecular dynamics is based on an extended Lagrangian \cite{Car1985}

\begin{align*}
\mathcal{L}&=
            \mu_\mathrm{e}\sum_i\int
            \left|\dot{\psi}_i(\mathbf{r})\right|^2\mathrm{d}\mathbf{r}
            +\frac{1}{2}\sum_I m_I\dot{\mathbf{R}}_I^2
            - E[\psi_i,\mathbf{R}]\\
            &+\sum_i\sum_j\Lambda_{ij}
            \left(\int \psi_i^\star(\mathbf{r})
            \psi_j(\mathbf{r})\mathrm{d}\mathbf{r}-\delta_{ij}\right)
\end{align*}

with the first two terms being the fictitious kinetic energy of the electrons and
the kinetic energy of the nuclei, respectively, and the last term is the constraints
resulting from the need to keep the orbitals orthogonal.

This results in the following equations of motion:

\begin{align*}
m_I\ddot{\mathbf{R}}_I&(t)=-\frac{\partial}{\partial\mathbf{R}_I}E[\psi_i,\mathbf{R}]
                          +\frac{\partial}{\partial\mathbf{R}_I}\;\left\{\text{constraints}\right\}\\
\mu_\mathrm{e}\ddot{\psi}_i&(t)=-\frac{\partial}{\partial\psi^\star_i}E[\psi_i,\mathbf{R}]
                      +\frac{\partial}{\partial\psi^\star_i}\;\left\{\text{constraints}\right\}
\end{align*}

The energy is calculated using the Kohn-Sham expression \cite{Hohenberg1964,Kohn1965}:

\begin{align*}
   E[\psi_i&,\mathbf{R}]=
   -\sum_i\frac{1}{2}\int \psi_i^\star(\mathbf{r})\nabla_i^2\psi_i(\mathbf{r})\mathrm{d}\mathbf{r}
   +\sum_{I<J}\frac{Z_IZ_J}{|\mathbf{R}_I-\mathbf{R}_J|}\\
   &-\sum_I\int\frac{Z_I\rho(\mathbf{r})}{|\mathbf{R}_I-\mathbf{r}|} \mathrm{d}\mathbf{r}
   +\frac{1}{2}\iint\frac{\rho(\mathbf{r})\rho(\mathbf{r'})}{|\mathbf{r}-\mathbf{r'}|}\mathrm{d}\mathbf{r}\mathrm{d}\mathbf{r'}\\
   &+ E_{\mathrm{xc}}[\psi_i]
   \qquad\quad\text{with}\;\rho(\mathbf{r})=\sum_i\int\psi_i^\star(\mathbf{r})\psi_i(\mathbf{r})\mathrm{d}\mathbf{r}
\end{align*}

We add the time-dependent external potential to the Kohn-Sham energy expression:

\begin{align*}
   E_t[\psi_i&,\mathbf{R}]=E[\psi_i,\mathbf{R}]+\\
            &\sum_I\int \frac{Z_Iv_\mathsf{sine}(\mathbf{r'},t)}{|\mathbf{R}_I-\mathbf{r'}|} \mathrm{d}\mathbf{r'}
                 -\iint\frac{v_\mathsf{sine}(\mathbf{r'},t)\rho(\mathbf{r})}{|\mathbf{r}-\mathbf{r'}|}\mathrm{d}\mathbf{r'}\mathrm{d}\mathbf{r}
\end{align*}

We have implemented into the Car-Parrinello molecular dynamics code \cite{cpmd}
a sinusoidal potential field which may change temporally and spatially
in three dimensions. For one spatial dimension:

\begin{equation*}
v_\mathsf{sine}(x,t)=\left(a(t)+A\right)\cdot\sin\left(2\bmpi\frac{x}{l_x}+b(t)\right)
\end{equation*}

$a(t)$ and $b(t)$ may be linear or sinusoidal functions.
With this flexible implementation, a broad range of phenomena can be simulated.
However, it must be considered that the wave length cannot exceed the size of the
unit cell as the spatial part $2\bmpi x / l_x$ ensures that the periodic boundary conditions are fulfilled.

\section{Applications}

\subsection{Electron Motion: Time-Dependent Amplitude}

Metal-molecule-metal junctions are used to measure the electron transport across 
single molecules.
Many experimental and theoretical studies aimed at understanding the mechanism of the electron transport (see \cite{Sun2006} and literature cited therein).
It was found that the resulting current depends on many factors which are difficult to
define experimentally. In particular the precise arrangement of a single molecule
between two surfaces or tips is not easy to control in experiment.
Most of the molecules under investigation are isolators and the distance between
the surfaces or tips is in the order of a nanometer, hence one may ask the question if
the effective current is significantly influenced by the linking molecules.
In an early study \cite{Joachim1995} of tunneling through a C$_{60}$ molecule
currents in the order of 10$^3$ nA were found at a voltage of 0.05 V which decrease by several orders of
magnitude if the tip-surface distance was increased above 10 $\AA$.
A better electronic connection is obtained if the molecule is covalently linked
to one or both surfaces. In a study of a self-assembled monolayer on a gold surface \cite{Donhauser2001}
a switching between an ON and an OFF state of organic molecules was observed.
It was characterized by the temporally changing apparent heights of the organic layer
(between 1 and 4 {\AA}) at an applied voltage of -1.4 V and a tunneling current in the order of 10$^{-4}$ nA.
The dependence of the tunneling current on covalent anchoring was studied in \cite{Selzer2002}.
Depending on the bonding situation and on the thickness of the organic layer,
current densities between 10$^{-7}$ and 10$^{-2}$ A/cm$^2$ were measured for voltages above 0.5 V.
Theoretical investigations were usually performed using perturbation theory approaches 
or studying the density of states obtained from density functional calculations \cite{Joachim2000,Piccinin2003,Sun2006}.
Direct dynamics calculations with an explicit external field give a more complete picture of the
full time evolution of a complex system.
In the present study we want to illustrate
the influence of a three-dimensional
electrical field on a metal-molecule-metal junction.
\\
We studied a system consisting of a dibenzenedithiol molecule covalently linked to two gold clusters.
The electric field is applied along the junction (Figure 1). \\ 

\begin{figure}
\includegraphics[height=2.2cm]{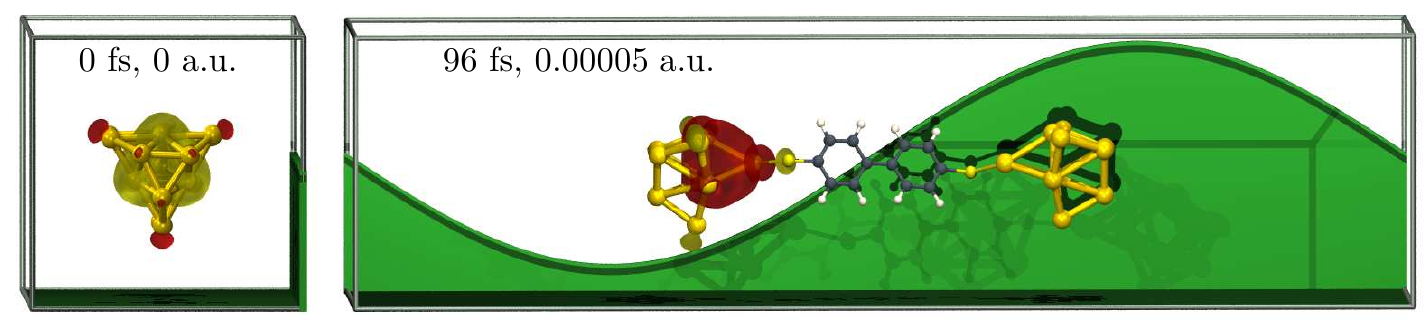}
\label{Fig1}
\caption{
Arrangement of the gold-dibenzenedithiol-gold junction in the supercell.
(grey:~C, white:~H, yellow:~S, gold:~Au)}
\end{figure}

A field with the following parameters was used:
A=0.000025 a.u.,
a(t)=$0.000025\cdot\sin\left(\frac{2\bmpi}{193.5\;\text{fs}}\cdot t-\frac{\bmpi}{2}\right)$ a.u.,
b(t)=0.

The amplitude varies sinusoidal with time between 0 and 0.00005~a.u. (0.03 kcal/mol, $\approx$ 1.4 kV/mm).
The transfer of one electron along the molecular junction is observed.

\begin{figure}
\includegraphics[height=12cm]{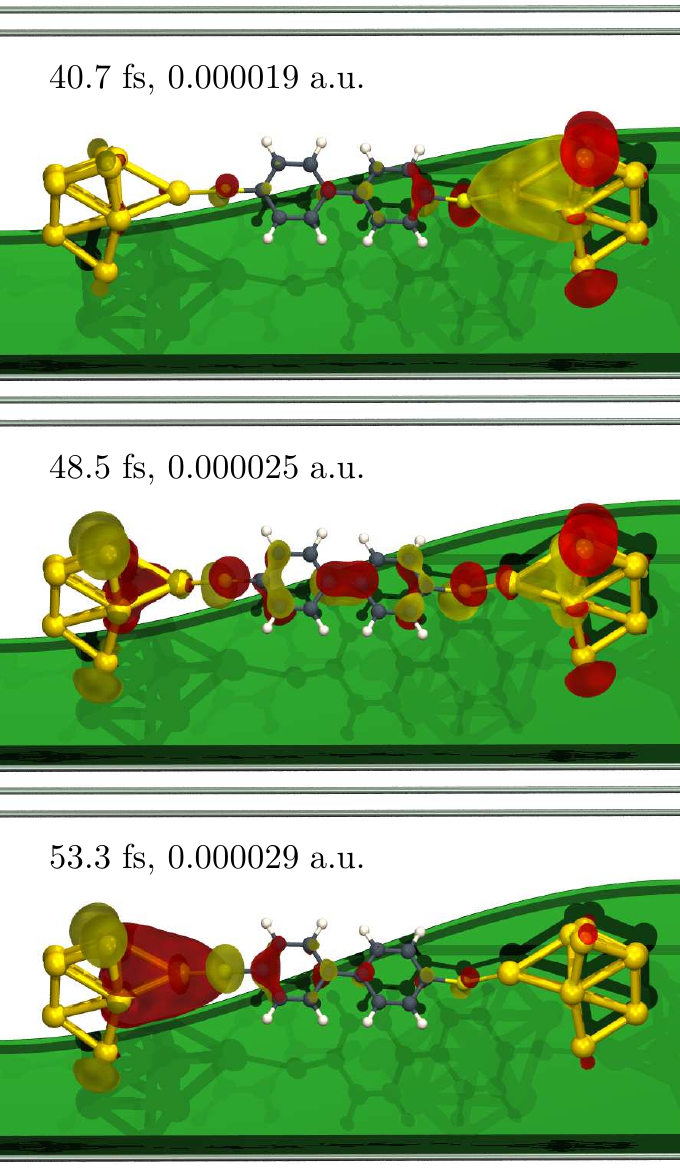}
\label{Fig2}
\caption{Snapshots of the motion of a localized orbital during the application
of an external field with increasing amplitude to a gold-dibenzenedithiol-gold junction.}
\end{figure}

The snapshots in Figure 2 show the motion of a localized (Wannier) orbital.
The orbital coefficients at the gold cluster at the right side of the figure decrease
and density is transfered to the aromatic system. Within about 10 fs the transfer to the
opposite cluster is completed.
Even if the motion of individual spin orbitals is observed, in total the spin effects cancel and the total
spin density (not shown) is essentially zero during the full process.
The external field causes a shift of the total charge rather than a build-up of spin density.
This shift occurs in a continuous motion. The total many-electron density is just slightly shifted 
to one side. 
 \\

\begin{figure}
\includegraphics[height=5cm]{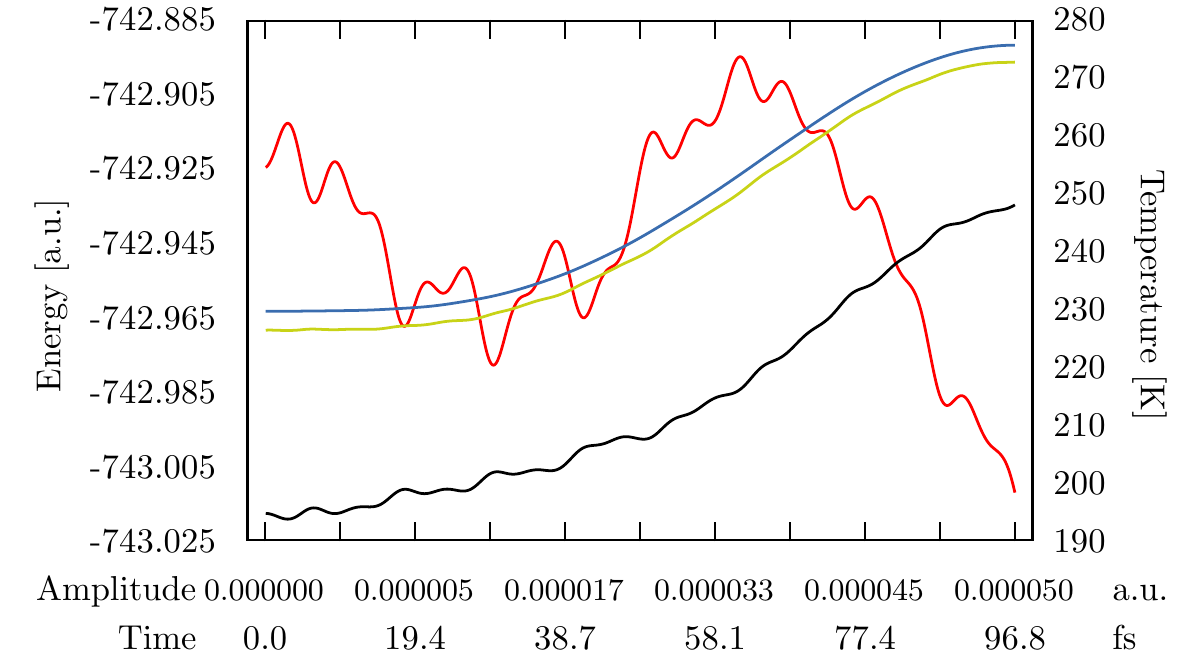}
\label{Fig3}
\caption{Total energy of the Car-Parrinello Lagrangian (blue), energy of the classical Hamiltonian
(green), Kohn-Sham energy (black), temperature (red) during the application of an external field
to a gold-dibenzenedithiol-gold junction.}
\end{figure}

Figure 3 shows the course of the energies and the temperature during the simulation run.
The total energy is raised by about 40 kcal/mol due to the application of the external field. \\

The electrons follow essentially instantaneously even if the fictitious mass of the electrons
is high in Car-Parrinello molecular dynamics simulations which results
in a slower electronic motion compared to experiment. The process is essentially adiabatic and is
determined by the change of the amplitude of the external field. These first results indicate
that the influence of the chemical nature of the junction on the current is small due to the
small metal-metal distance. However, the intermediate build-up of
orbital density in the aromatic system shows that it 
is different from zero, the aromatic system takes actively part in the charge transfer.

\subsection{Ion Migration: Time-Dependent Phase}

$\boldsymbol{\beta}$-Eucryptite (LiAlSiO$_{\text{4}}$) \cite{Alpen1977} is known as a one-dimensional  Li$^+$ ionic conductor, a substance class which is of high interest for the development of batteries \cite{Murugan2007,Buschmann2011}.
For a similar material, Li$_7$La$_3$Zr$_2$O$_{12}$, an unusual concerted mechanism for ion migration has been found 
in a recent theoretical study using
first principles equilibrium simulations at high temperatures \cite{Jalem2013}.
In $\boldsymbol{\beta}$-eucryptite the conductivity is along the crystallographic c-axis of the quartz-like structure.
To study the migration of the ions in the crystal,
a system is modeled with one crystal defect introduced in the supercell (Figure 4).
 A Li$^+$ ion is removed and a Si$^{4+}$ ion is substituted by an Al$^{3+}$ ion to obtain a neutral system.
 The potential field is applied along the c-axis.
 The phase of the potential field varies linearly in time.\\ 

\begin{figure}
\includegraphics[height=5cm]{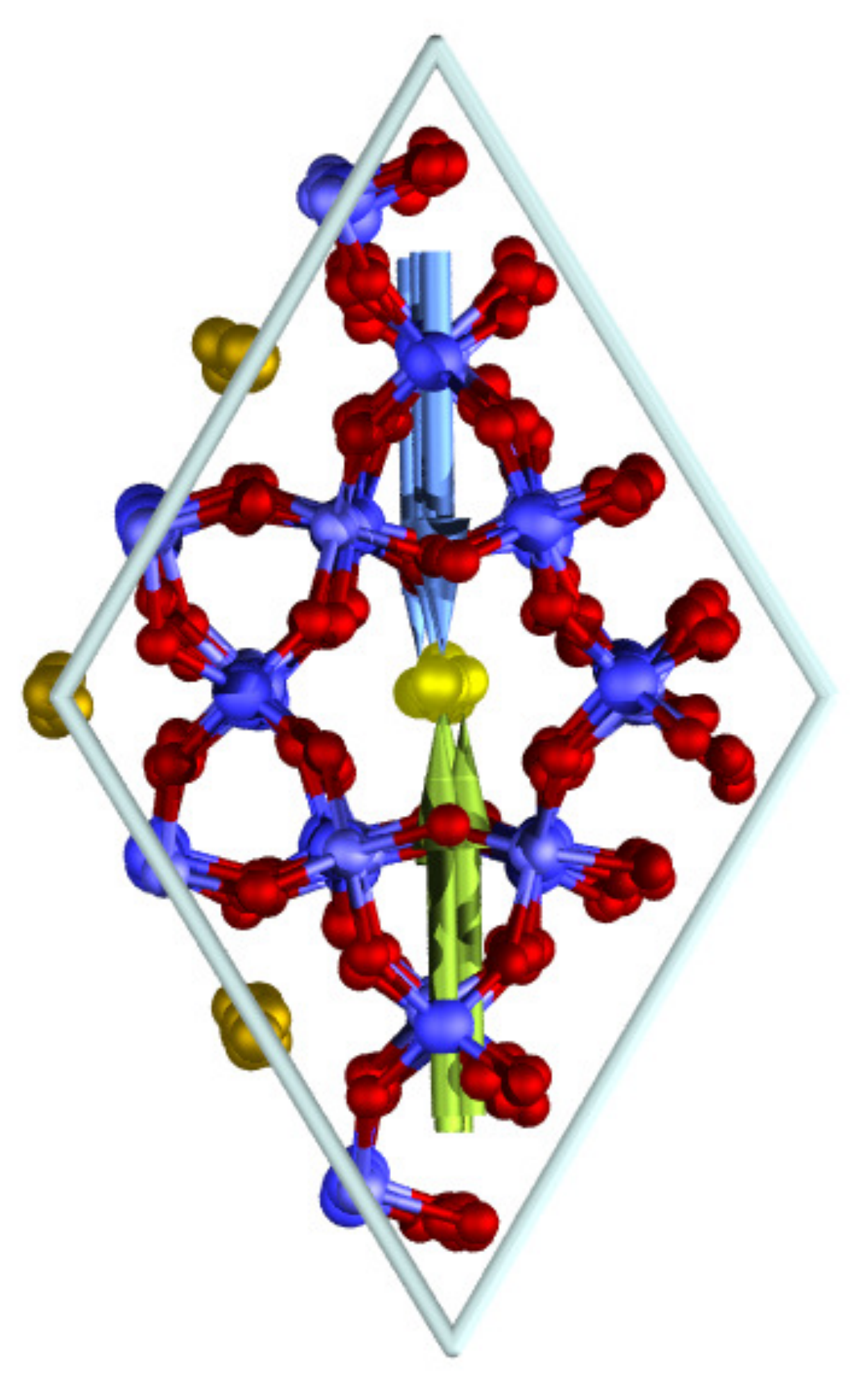}
\label{Fig4}
\caption{
Unit cell of $\boldsymbol{\beta}$-eucryptite. Within the unit cell the lithium ions may migrate along four different 
channels. Dark blue: Si, blue: Al, red: O, gold: Li, yellow: Li channel with vacancy. Only in this channel Li motion
is observed.
}
\end{figure}

A field with the following parameters was used:
A=0.0001 a.u.,
a(t)=0,
b(t)=$\frac{2\bmpi}{483.8\;\text{fs}}\cdot t$.
The field which has an amplitude of 0.0001 a.u. (0.06 kcal/mol, $\approx$ 3.6 kV/mm)
is moving with a velocity of 15000 m/s, respectively.
This corresponds to a phase shift of 2$\pi$ within 0.29 ps.
The series of snapshots in (Figure 5) shows the migration of five Li$^+$-ions
to new lattice sites on a time scale of a few hundred femtoseconds as indicated by the moving arrows. The motion
is started by a lithium ion near the left border of the simulation cell.
The neighbouring lithium ions follow till the original vacancy is filled and
a new vacancy near the left border of the simulation cell is generated. On a
longer time scale this vacancy would be filled by lithium ions from the
right border (respecting periodic boundary conditions). \\

\begin{figure}
\includegraphics[height=6.4cm]{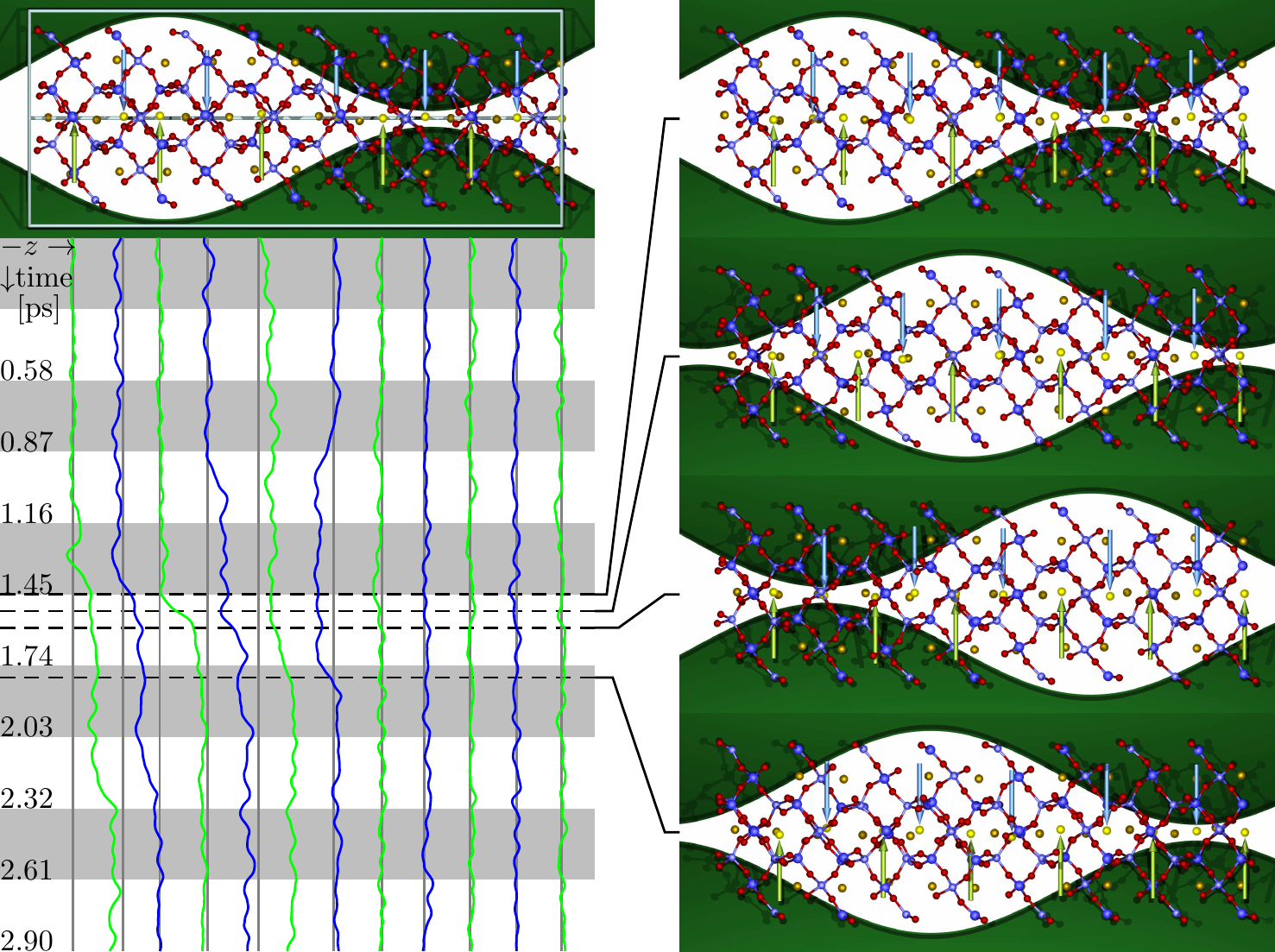}
\label{Fig5}
\caption{
Snapshots from a CPMD simulation showing the motion of lithium ions (yellow) in $\boldsymbol{\beta}$-eucryptite.
The channel containing a vacancy is shown in light yellow. Dark blue: Si, blue: Al, red: O.
The arrows mark the migrating lithium ions. The migration can also be followed from the
trajectories to the left. Five lithium ions in the left half of the simulation cell shown
migrate to the next lattice site generating a new vacancy at the left border of the simulation cell.
}
\end{figure}

In Figure 6 the temporal evolution of the energies and temperature is shown.
The total energy shows a slight oscillation in the beginning when the field is turned on.
The kinetic energy of the electrons
(difference between the total energy and the energy of the classical Hamiltonian) stay
small. The temperature (corresponding to the difference between the energy of the
classical Hamiltonian and the Kohn-Sham energy) shows a slight increase by about
30 K. The ion migrations are possible with a 
rather moderate take-up of energy.

\begin{figure}
\includegraphics[height=5cm]{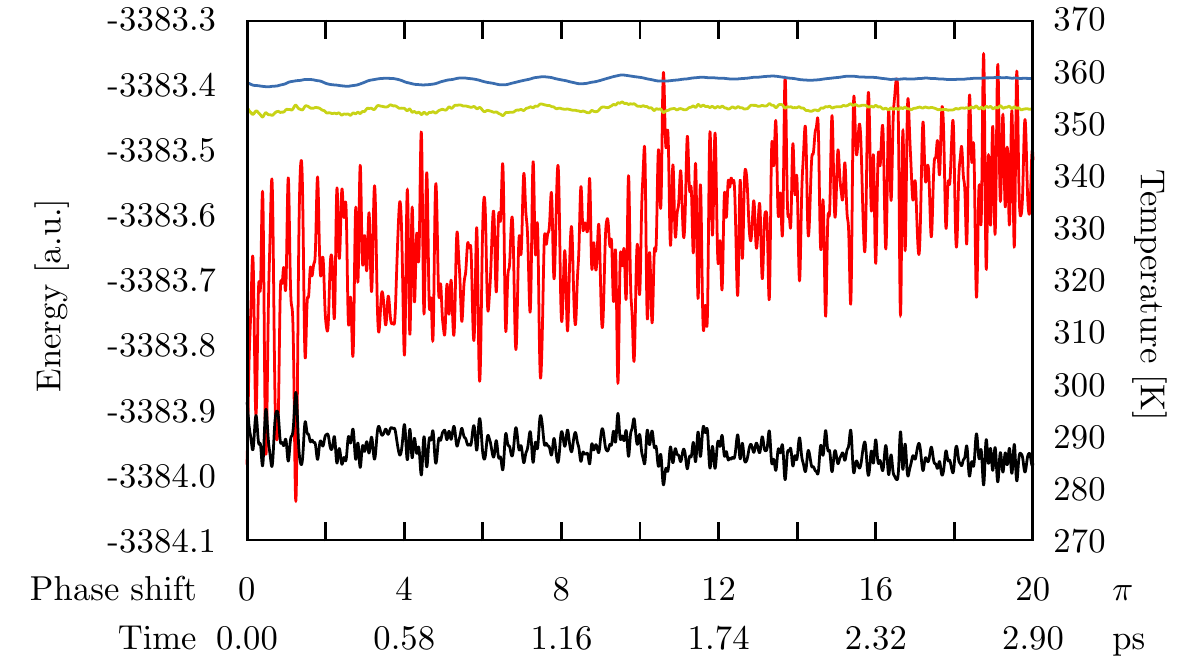}
\label{Fig6}
\caption{Total energy of the Car-Parrinello Lagrangian (blue), energy of the classical Hamiltonian
(green), Kohn-Sham energy (black), temperature (red) during the application of a temporally
moving external field to $\boldsymbol{\beta}$-eucryptite.
}
\end{figure}

\section{Conclusions}

We have added a sinusoidal field to Car-Parrinello molecular dynamics.
The field can change with phase and amplitude. Like this it is possible to
model a vast diversity of experimental situations in which an electrical field
matters. 
In principle, the approach can be extended to more complex periodic fields by using more than one
sine function. 
In first applications we have illustrated the application to electron and ion transfers.
For a metal-molecule-metal junction we find that the electrons
follow the electric field nearly adiabatically along the junction.
For a lithium ion conductor we find a consecutive hopping of ions to neighbouring lattice
sites along a channel containing a defect.
The method is quantitative in principle, however, 
the simulations are limited to periodic fields with wave lengths not larger than the unit cell lengths 
and to time scales which allow for very fast changing fields only.
Being able to follow the complex electronic and ionic motion is the main advancement from such first-principle calculations.
The simulations may serve as a check if simpler models are applicable.

\section{Methods}
The simulations were performed using Car-Parrinello molecular dynamics
\cite{Car1985} as implemented in the CPMD plane wave code
\cite{cpmd}. 
For all calculations the BLYP exchange-correlation
functional \cite{Becke1988,Lee1988} was used in its unrestricted formulation (LSD).
Troullier-Martins pseudopotentials were
employed for describing the core electrons \cite{Troullier1993}.
The pseudopotential cutoff was set to 70.0~Rydberg.
The fictitious electron mass was set to 400 a.u. and a time step of 4 a.u. (0.097 fs) was
used.  After the equilibration, the temperature was not controlled. \\
For the simulations of the metal-molecule-metal junctions a periodically repeated simulation cell
with a size of 15 x 15 x 60~\AA$^3$ was used.
The systems were initially equilibrated at a temperature of 150 K in order to reduce molecular vibrations
and to focus on the electronic motion. \\
For the simulations of the eucryptite crystal periodic boundary conditions were applied
with a cell size of 10.5 x 9.1 x 44.8~\AA$^3$. The systems were equilibrated at 300 K.

\section{Acknowledgements}
We thank Marius Schulte for helpful discussions.

\bibliography{../../literature.bib}

\begin{thebibliography}{23}%
\makeatletter
\providecommand \@ifxundefined [1]{%
 \@ifx{#1\undefined}
}%
\providecommand \@ifnum [1]{%
 \ifnum #1\expandafter \@firstoftwo
 \else \expandafter \@secondoftwo
 \fi
}%
\providecommand \@ifx [1]{%
 \ifx #1\expandafter \@firstoftwo
 \else \expandafter \@secondoftwo
 \fi
}%
\providecommand \natexlab [1]{#1}%
\providecommand \enquote  [1]{``#1''}%
\providecommand \bibnamefont  [1]{#1}%
\providecommand \bibfnamefont [1]{#1}%
\providecommand \citenamefont [1]{#1}%
\providecommand \href@noop [0]{\@secondoftwo}%
\providecommand \href [0]{\begingroup \@sanitize@url \@href}%
\providecommand \@href[1]{\@@startlink{#1}\@@href}%
\providecommand \@@href[1]{\endgroup#1\@@endlink}%
\providecommand \@sanitize@url [0]{\catcode `\\12\catcode `\$12\catcode
  `\&12\catcode `\#12\catcode `\^12\catcode `\_12\catcode `\%12\relax}%
\providecommand \@@startlink[1]{}%
\providecommand \@@endlink[0]{}%
\providecommand \url  [0]{\begingroup\@sanitize@url \@url }%
\providecommand \@url [1]{\endgroup\@href {#1}{\urlprefix }}%
\providecommand \urlprefix  [0]{URL }%
\providecommand \Eprint [0]{\href }%
\providecommand \doibase [0]{http://dx.doi.org/}%
\providecommand \selectlanguage [0]{\@gobble}%
\providecommand \bibinfo  [0]{\@secondoftwo}%
\providecommand \bibfield  [0]{\@secondoftwo}%
\providecommand \translation [1]{[#1]}%
\providecommand \BibitemOpen [0]{}%
\providecommand \bibitemStop [0]{}%
\providecommand \bibitemNoStop [0]{.\EOS\space}%
\providecommand \EOS [0]{\spacefactor3000\relax}%
\providecommand \BibitemShut  [1]{\csname bibitem#1\endcsname}%
\let\auto@bib@innerbib\@empty
\bibitem [{\citenamefont {Car}\ and\ \citenamefont
  {Parrinello}(1985)}]{Car1985}%
  \BibitemOpen
  \bibfield  {author} {\bibinfo {author} {\bibfnamefont {R.}~\bibnamefont
  {Car}}\ and\ \bibinfo {author} {\bibfnamefont {M.}~\bibnamefont
  {Parrinello}},\ }\href@noop {} {\bibfield  {journal} {\bibinfo  {journal}
  {Phys. Rev. Lett.}\ }\textbf {\bibinfo {volume} {55}},\ \bibinfo {pages}
  {2471} (\bibinfo {year} {1985})}\BibitemShut {NoStop}%
\bibitem [{CPMD()}]{cpmd}%
  \BibitemOpen
  CPMD,\ \href@noop {} {}\bibinfo {note} {{CPMD}, Version 3.15, {J}. {H}utter
  et al., http://www.cpmd.org/, Copyright {IBM} Corp 1990-2008, Copyright {MPI}
  f{\"u}r {F}estk{\"o}rperforschung {S}tuttgart 1997-2001}\BibitemShut
  {NoStop}%
\bibitem [{\citenamefont {Hohenberg}\ and\ \citenamefont
  {Kohn}(1964)}]{Hohenberg1964}%
  \BibitemOpen
  \bibfield  {author} {\bibinfo {author} {\bibfnamefont {P.}~\bibnamefont
  {Hohenberg}}\ and\ \bibinfo {author} {\bibfnamefont {W.}~\bibnamefont
  {Kohn}},\ }\href@noop {} {\bibfield  {journal} {\bibinfo  {journal} {Phys.
  Rev. B}\ }\textbf {\bibinfo {volume} {136}},\ \bibinfo {pages} {864}
  (\bibinfo {year} {1964})}\BibitemShut {NoStop}%
\bibitem [{\citenamefont {Kohn}\ and\ \citenamefont {Sham}(1965)}]{Kohn1965}%
  \BibitemOpen
  \bibfield  {author} {\bibinfo {author} {\bibfnamefont {W.}~\bibnamefont
  {Kohn}}\ and\ \bibinfo {author} {\bibfnamefont {L.~J.}\ \bibnamefont
  {Sham}},\ }\href@noop {} {\bibfield  {journal} {\bibinfo  {journal} {Phys.
  Rev. A}\ }\textbf {\bibinfo {volume} {140}},\ \bibinfo {pages} {1133}
  (\bibinfo {year} {1965})}\BibitemShut {NoStop}%
\bibitem [{\citenamefont {Kudin}\ \emph
  {et~al.}(2007{\natexlab{a}})\citenamefont {Kudin}, \citenamefont {Car},\ and\
  \citenamefont {Resta}}]{Kudin2007a}%
  \BibitemOpen
  \bibfield  {author} {\bibinfo {author} {\bibfnamefont {K.~N.}\ \bibnamefont
  {Kudin}}, \bibinfo {author} {\bibfnamefont {R.}~\bibnamefont {Car}}, \ and\
  \bibinfo {author} {\bibfnamefont {R.}~\bibnamefont {Resta}},\ }\href@noop {}
  {\bibfield  {journal} {\bibinfo  {journal} {J. Chem. Phys.}\ }\textbf
  {\bibinfo {volume} {126}},\ \bibinfo {pages} {234101} (\bibinfo {year}
  {2007}{\natexlab{a}})}\BibitemShut {NoStop}%
\bibitem [{\citenamefont {Kudin}\ \emph
  {et~al.}(2007{\natexlab{b}})\citenamefont {Kudin}, \citenamefont {Car},\ and\
  \citenamefont {Resta}}]{Kudin2007b}%
  \BibitemOpen
  \bibfield  {author} {\bibinfo {author} {\bibfnamefont {K.~N.}\ \bibnamefont
  {Kudin}}, \bibinfo {author} {\bibfnamefont {R.}~\bibnamefont {Car}}, \ and\
  \bibinfo {author} {\bibfnamefont {R.}~\bibnamefont {Resta}},\ }\href@noop {}
  {\bibfield  {journal} {\bibinfo  {journal} {J. Chem. Phys.}\ }\textbf
  {\bibinfo {volume} {127}},\ \bibinfo {pages} {194902} (\bibinfo {year}
  {2007}{\natexlab{b}})}\BibitemShut {NoStop}%
\bibitem [{\citenamefont {Springborg}\ and\ \citenamefont
  {Kirtman}(2007)}]{Springborg2007}%
  \BibitemOpen
  \bibfield  {author} {\bibinfo {author} {\bibfnamefont {M.}~\bibnamefont
  {Springborg}}\ and\ \bibinfo {author} {\bibfnamefont {B.}~\bibnamefont
  {Kirtman}},\ }\href@noop {} {\bibfield  {journal} {\bibinfo  {journal} {J.
  Chem. Phys.}\ }\textbf {\bibinfo {volume} {126}},\ \bibinfo {pages} {104107}
  (\bibinfo {year} {2007})}\BibitemShut {NoStop}%
\bibitem [{\citenamefont {Springborg}\ and\ \citenamefont
  {Kirtman}(2008{\natexlab{a}})}]{Springborg2008a}%
  \BibitemOpen
  \bibfield  {author} {\bibinfo {author} {\bibfnamefont {M.}~\bibnamefont
  {Springborg}}\ and\ \bibinfo {author} {\bibfnamefont {B.}~\bibnamefont
  {Kirtman}},\ }\href@noop {} {\bibfield  {journal} {\bibinfo  {journal} {Phys.
  Rev. B}\ }\textbf {\bibinfo {volume} {77}},\ \bibinfo {pages} {045102}
  (\bibinfo {year} {2008}{\natexlab{a}})}\BibitemShut {NoStop}%
\bibitem [{\citenamefont {Springborg}\ and\ \citenamefont
  {Kirtman}(2008{\natexlab{b}})}]{Springborg2008b}%
  \BibitemOpen
  \bibfield  {author} {\bibinfo {author} {\bibfnamefont {M.}~\bibnamefont
  {Springborg}}\ and\ \bibinfo {author} {\bibfnamefont {B.}~\bibnamefont
  {Kirtman}},\ }\href@noop {} {\bibfield  {journal} {\bibinfo  {journal} {Phys.
  Rev. E}\ }\textbf {\bibinfo {volume} {77}},\ \bibinfo {pages} {209901}
  (\bibinfo {year} {2008}{\natexlab{b}})}\BibitemShut {NoStop}%
\bibitem [{\citenamefont {Kirtman}\ \emph {et~al.}(2009)\citenamefont
  {Kirtman}, \citenamefont {Ferrero}, \citenamefont {R{\'e}rat},\ and\
  \citenamefont {Springborg}}]{Kirtman2009}%
  \BibitemOpen
  \bibfield  {author} {\bibinfo {author} {\bibfnamefont {B.}~\bibnamefont
  {Kirtman}}, \bibinfo {author} {\bibfnamefont {M.}~\bibnamefont {Ferrero}},
  \bibinfo {author} {\bibfnamefont {M.}~\bibnamefont {R{\'e}rat}}, \ and\
  \bibinfo {author} {\bibfnamefont {M.}~\bibnamefont {Springborg}},\
  }\href@noop {} {\bibfield  {journal} {\bibinfo  {journal} {J. Chem. Phys.}\
  }\textbf {\bibinfo {volume} {131}},\ \bibinfo {pages} {044109} (\bibinfo
  {year} {2009})}\BibitemShut {NoStop}%
\bibitem [{\citenamefont {Sun}\ \emph {et~al.}(2006)\citenamefont {Sun},
  \citenamefont {Selloni},\ and\ \citenamefont {Scoles}}]{Sun2006}%
  \BibitemOpen
  \bibfield  {author} {\bibinfo {author} {\bibfnamefont {Q.}~\bibnamefont
  {Sun}}, \bibinfo {author} {\bibfnamefont {A.}~\bibnamefont {Selloni}}, \ and\
  \bibinfo {author} {\bibfnamefont {G.}~\bibnamefont {Scoles}},\ }\href@noop {}
  {\bibfield  {journal} {\bibinfo  {journal} {J. Phys. Chem. B}\ }\textbf
  {\bibinfo {volume} {110}},\ \bibinfo {pages} {3493} (\bibinfo {year}
  {2006})}\BibitemShut {NoStop}%
\bibitem [{\citenamefont {Joachim}\ \emph {et~al.}(1995)\citenamefont
  {Joachim}, \citenamefont {Gimzewski}, \citenamefont {Schlittler},\ and\
  \citenamefont {Chavy}}]{Joachim1995}%
  \BibitemOpen
  \bibfield  {author} {\bibinfo {author} {\bibfnamefont {C.}~\bibnamefont
  {Joachim}}, \bibinfo {author} {\bibfnamefont {J.~K.}\ \bibnamefont
  {Gimzewski}}, \bibinfo {author} {\bibfnamefont {R.~R.}\ \bibnamefont
  {Schlittler}}, \ and\ \bibinfo {author} {\bibfnamefont {C.}~\bibnamefont
  {Chavy}},\ }\href@noop {} {\bibfield  {journal} {\bibinfo  {journal} {Phys.
  Rev. Lett.}\ }\textbf {\bibinfo {volume} {74}},\ \bibinfo {pages} {2102}
  (\bibinfo {year} {1995})}\BibitemShut {NoStop}%
\bibitem [{\citenamefont {Donhauser}\ \emph {et~al.}(2001)\citenamefont
  {Donhauser}, \citenamefont {Mantooth}, \citenamefont {Kelly}, \citenamefont
  {Bumm}, \citenamefont {Monell}, \citenamefont {Stapleton}, \citenamefont
  {Price}, \citenamefont {Rawlett}, \citenamefont {Allara}, \citenamefont
  {Tour},\ and\ \citenamefont {Weiss}}]{Donhauser2001}%
  \BibitemOpen
  \bibfield  {author} {\bibinfo {author} {\bibfnamefont {Z.~J.}\ \bibnamefont
  {Donhauser}}, \bibinfo {author} {\bibfnamefont {B.~A.}\ \bibnamefont
  {Mantooth}}, \bibinfo {author} {\bibfnamefont {K.~F.}\ \bibnamefont {Kelly}},
  \bibinfo {author} {\bibfnamefont {L.~A.}\ \bibnamefont {Bumm}}, \bibinfo
  {author} {\bibfnamefont {J.~D.}\ \bibnamefont {Monell}}, \bibinfo {author}
  {\bibfnamefont {J.~J.}\ \bibnamefont {Stapleton}}, \bibinfo {author}
  {\bibfnamefont {D.~W.}\ \bibnamefont {Price}}, \bibinfo {author}
  {\bibfnamefont {A.~M.}\ \bibnamefont {Rawlett}}, \bibinfo {author}
  {\bibfnamefont {D.~L.}\ \bibnamefont {Allara}}, \bibinfo {author}
  {\bibfnamefont {J.~M.}\ \bibnamefont {Tour}}, \ and\ \bibinfo {author}
  {\bibfnamefont {P.~S.}\ \bibnamefont {Weiss}},\ }\href@noop {} {\bibfield
  {journal} {\bibinfo  {journal} {Science}\ }\textbf {\bibinfo {volume}
  {292}},\ \bibinfo {pages} {2303} (\bibinfo {year} {2001})}\BibitemShut
  {NoStop}%
\bibitem [{\citenamefont {Selzer}\ \emph {et~al.}(2002)\citenamefont {Selzer},
  \citenamefont {Salomon},\ and\ \citenamefont {Cahen}}]{Selzer2002}%
  \BibitemOpen
  \bibfield  {author} {\bibinfo {author} {\bibfnamefont {Y.}~\bibnamefont
  {Selzer}}, \bibinfo {author} {\bibfnamefont {A.}~\bibnamefont {Salomon}}, \
  and\ \bibinfo {author} {\bibfnamefont {D.}~\bibnamefont {Cahen}},\
  }\href@noop {} {\bibfield  {journal} {\bibinfo  {journal} {J. Phys. Chem. B}\
  }\textbf {\bibinfo {volume} {106}},\ \bibinfo {pages} {10432} (\bibinfo
  {year} {2002})}\BibitemShut {NoStop}%
\bibitem [{\citenamefont {Joachim}\ \emph {et~al.}(2000)\citenamefont
  {Joachim}, \citenamefont {Gimzewski},\ and\ \citenamefont
  {Aviram}}]{Joachim2000}%
  \BibitemOpen
  \bibfield  {author} {\bibinfo {author} {\bibfnamefont {C.}~\bibnamefont
  {Joachim}}, \bibinfo {author} {\bibfnamefont {J.~K.}\ \bibnamefont
  {Gimzewski}}, \ and\ \bibinfo {author} {\bibfnamefont {A.}~\bibnamefont
  {Aviram}},\ }\href@noop {} {\bibfield  {journal} {\bibinfo  {journal}
  {Nature}\ }\textbf {\bibinfo {volume} {408}},\ \bibinfo {pages} {541}
  (\bibinfo {year} {2000})}\BibitemShut {NoStop}%
\bibitem [{\citenamefont {Piccinin}\ \emph {et~al.}(2003)\citenamefont
  {Piccinin}, \citenamefont {Selloni}, \citenamefont {Scandolo}, \citenamefont
  {Car},\ and\ \citenamefont {Scoles}}]{Piccinin2003}%
  \BibitemOpen
  \bibfield  {author} {\bibinfo {author} {\bibfnamefont {S.}~\bibnamefont
  {Piccinin}}, \bibinfo {author} {\bibfnamefont {A.}~\bibnamefont {Selloni}},
  \bibinfo {author} {\bibfnamefont {S.}~\bibnamefont {Scandolo}}, \bibinfo
  {author} {\bibfnamefont {R.}~\bibnamefont {Car}}, \ and\ \bibinfo {author}
  {\bibfnamefont {G.}~\bibnamefont {Scoles}},\ }\href@noop {} {\bibfield
  {journal} {\bibinfo  {journal} {J. Chem. Phys.}\ }\textbf {\bibinfo {volume}
  {119}},\ \bibinfo {pages} {6729} (\bibinfo {year} {2003})}\BibitemShut
  {NoStop}%
\bibitem [{\citenamefont {Alpen}\ \emph {et~al.}(1977)\citenamefont {Alpen},
  \citenamefont {Sch{\"o}nherr}, \citenamefont {Schulz},\ and\ \citenamefont
  {Talat}}]{Alpen1977}%
  \BibitemOpen
  \bibfield  {author} {\bibinfo {author} {\bibfnamefont {U.~V.}\ \bibnamefont
  {Alpen}}, \bibinfo {author} {\bibfnamefont {E.}~\bibnamefont
  {Sch{\"o}nherr}}, \bibinfo {author} {\bibfnamefont {H.}~\bibnamefont
  {Schulz}}, \ and\ \bibinfo {author} {\bibfnamefont {G.~H.}\ \bibnamefont
  {Talat}},\ }\href@noop {} {\bibfield  {journal} {\bibinfo  {journal}
  {Electrochimica Acta}\ }\textbf {\bibinfo {volume} {22}},\ \bibinfo {pages}
  {805} (\bibinfo {year} {1977})}\BibitemShut {NoStop}%
\bibitem [{\citenamefont {Murugan}\ \emph {et~al.}(2007)\citenamefont
  {Murugan}, \citenamefont {Thangadurai},\ and\ \citenamefont
  {Weppner}}]{Murugan2007}%
  \BibitemOpen
  \bibfield  {author} {\bibinfo {author} {\bibfnamefont {R.}~\bibnamefont
  {Murugan}}, \bibinfo {author} {\bibfnamefont {V.}~\bibnamefont
  {Thangadurai}}, \ and\ \bibinfo {author} {\bibfnamefont {W.}~\bibnamefont
  {Weppner}},\ }\href@noop {} {\bibfield  {journal} {\bibinfo  {journal}
  {Angew. Chem.}\ }\textbf {\bibinfo {volume} {46}},\ \bibinfo {pages} {7778}
  (\bibinfo {year} {2007})}\BibitemShut {NoStop}%
\bibitem [{\citenamefont {Buschmann}\ \emph {et~al.}(2011)\citenamefont
  {Buschmann}, \citenamefont {D{\"o}lle}, \citenamefont {Berendts},
  \citenamefont {Kuhn}, \citenamefont {Bottke}, \citenamefont {Wilkening},
  \citenamefont {Heitjans}, \citenamefont {Senyshyn}, \citenamefont
  {Ehrenberg}, \citenamefont {Lotnyk}, \citenamefont {Duppel}, \citenamefont
  {Kienle},\ and\ \citenamefont {Janek}}]{Buschmann2011}%
  \BibitemOpen
  \bibfield  {author} {\bibinfo {author} {\bibfnamefont {H.}~\bibnamefont
  {Buschmann}}, \bibinfo {author} {\bibfnamefont {J.}~\bibnamefont
  {D{\"o}lle}}, \bibinfo {author} {\bibfnamefont {S.}~\bibnamefont {Berendts}},
  \bibinfo {author} {\bibfnamefont {A.}~\bibnamefont {Kuhn}}, \bibinfo {author}
  {\bibfnamefont {P.}~\bibnamefont {Bottke}}, \bibinfo {author} {\bibfnamefont
  {M.}~\bibnamefont {Wilkening}}, \bibinfo {author} {\bibfnamefont
  {P.}~\bibnamefont {Heitjans}}, \bibinfo {author} {\bibfnamefont
  {A.}~\bibnamefont {Senyshyn}}, \bibinfo {author} {\bibfnamefont
  {H.}~\bibnamefont {Ehrenberg}}, \bibinfo {author} {\bibfnamefont
  {A.}~\bibnamefont {Lotnyk}}, \bibinfo {author} {\bibfnamefont
  {V.}~\bibnamefont {Duppel}}, \bibinfo {author} {\bibfnamefont
  {L.}~\bibnamefont {Kienle}}, \ and\ \bibinfo {author} {\bibfnamefont
  {J.}~\bibnamefont {Janek}},\ }\href@noop {} {\bibfield  {journal} {\bibinfo
  {journal} {Phys. Chem. Chem. Phys.}\ }\textbf {\bibinfo {volume} {13}},\
  \bibinfo {pages} {19378} (\bibinfo {year} {2011})}\BibitemShut {NoStop}%
\bibitem [{\citenamefont {Jalem}\ \emph {et~al.}(2013)\citenamefont {Jalem},
  \citenamefont {Yamamoto}, \citenamefont {Shiiba}, \citenamefont {Nakayama},
  \citenamefont {Munakata}, \citenamefont {Kasuga},\ and\ \citenamefont
  {Kanamura}}]{Jalem2013}%
  \BibitemOpen
  \bibfield  {author} {\bibinfo {author} {\bibfnamefont {R.}~\bibnamefont
  {Jalem}}, \bibinfo {author} {\bibfnamefont {Y.}~\bibnamefont {Yamamoto}},
  \bibinfo {author} {\bibfnamefont {H.}~\bibnamefont {Shiiba}}, \bibinfo
  {author} {\bibfnamefont {M.}~\bibnamefont {Nakayama}}, \bibinfo {author}
  {\bibfnamefont {H.}~\bibnamefont {Munakata}}, \bibinfo {author}
  {\bibfnamefont {T.}~\bibnamefont {Kasuga}}, \ and\ \bibinfo {author}
  {\bibfnamefont {K.}~\bibnamefont {Kanamura}},\ }\href@noop {} {\bibfield
  {journal} {\bibinfo  {journal} {Chem. Mater.}\ }\textbf {\bibinfo {volume}
  {25}},\ \bibinfo {pages} {425} (\bibinfo {year} {2013})}\BibitemShut
  {NoStop}%
\bibitem [{\citenamefont {Becke}(1988)}]{Becke1988}%
  \BibitemOpen
  \bibfield  {author} {\bibinfo {author} {\bibfnamefont {A.}~\bibnamefont
  {Becke}},\ }\href@noop {} {\bibfield  {journal} {\bibinfo  {journal} {Phys.
  Rev. A}\ }\textbf {\bibinfo {volume} {38}},\ \bibinfo {pages} {3098}
  (\bibinfo {year} {1988})}\BibitemShut {NoStop}%
\bibitem [{\citenamefont {Lee}\ \emph {et~al.}(1988)\citenamefont {Lee},
  \citenamefont {Yang},\ and\ \citenamefont {Parr}}]{Lee1988}%
  \BibitemOpen
  \bibfield  {author} {\bibinfo {author} {\bibfnamefont {C.}~\bibnamefont
  {Lee}}, \bibinfo {author} {\bibfnamefont {W.}~\bibnamefont {Yang}}, \ and\
  \bibinfo {author} {\bibfnamefont {R.~G.}\ \bibnamefont {Parr}},\ }\href@noop
  {} {\bibfield  {journal} {\bibinfo  {journal} {Phys. Rev. B}\ }\textbf
  {\bibinfo {volume} {37}},\ \bibinfo {pages} {785} (\bibinfo {year}
  {1988})}\BibitemShut {NoStop}%
\bibitem [{\citenamefont {Troullier}\ and\ \citenamefont
  {Martins}(1991)}]{Troullier1993}%
  \BibitemOpen
  \bibfield  {author} {\bibinfo {author} {\bibfnamefont {N.}~\bibnamefont
  {Troullier}}\ and\ \bibinfo {author} {\bibfnamefont {J.~L.}\ \bibnamefont
  {Martins}},\ }\href@noop {} {\bibfield  {journal} {\bibinfo  {journal} {Phys.
  Rev.}\ }\textbf {\bibinfo {volume} {43}},\ \bibinfo {pages} {1993} (\bibinfo
  {year} {1991})}\BibitemShut {NoStop}%
\end{thebibliography}%

\end{document}